\documentclass{article}
\usepackage{graphics}
\usepackage{graphicx}
\usepackage{amssymb}
\usepackage{array}
\usepackage{fancyhdr}
\usepackage{subfigure}
\usepackage{graphicx}
\usepackage{mathrsfs}
\usepackage{slashed}
\usepackage{stmaryrd}
\usepackage{epsfig}
\usepackage{dcolumn}
\usepackage{bm}
\usepackage{amsmath}
\usepackage{wrapfig}
\usepackage{makeidx}
\usepackage{wasysym}
 \usepackage{relsize}
\usepackage{la}
\usepackage[utf8]{inputenc}

\usepackage{array}
\usepackage{fancyhdr}
\usepackage{graphicx}
\usepackage{mathrsfs}
\usepackage{slashed}
\usepackage{stmaryrd}
\begin{document}
\title{\bf Thermalized Vacuum and Vacuum Effects}
\author{Yi-Cheng Huang\\
E-mail: {\tt ychuang1109@msn.com}
}                     
%
%
\date{}
\maketitle
%
%
{\bf Abstract}
\vskip.2cm

{
Some of the well-known effects regarding the vacuum are revisited under the formalism of the imaginary-time field theory. From these effects, they could imply the existence of one thermal vacuum in different circumstances. The imaginary-time hamiltonian of the vacuum is found to provide not only exact distribution functions in the calculations of the Casimir effect and the van der Waals force but also cutoff functions.   The thermal bath for the Unruh effect is constructed from the imaginary-time Green function. From the field theory in the curved space-time,  field quantizations are defined according to different vacuum states  and lead to the Hawking radiation; the introduced conformal invariance agrees with the formalism of the imaginary-time field theory. The induced Green functions in the curved space-time are in accordance with  those from the picture given from the thermal vacuum.

 %
} 
%
%
\newpage
\section{Introduction}
\label{intro}
In the previous work \cite{huang13a}, a thermal theory of virtual particles in the imaginary-time and space has been constructed and its applications on the calculations of one-loop radiative corrections and the corresponding renormalization group equations are performed. The conclusions show that a vacuum with a small temperature could generate consistent results with those in the successful QED, at least in the one-loop level so far. As a further discussion of this imaginary-time theory, the goal of this paper is to relate it with the known effects of vacuum and to attempt describing them within one universal picture.
There have been lots of discussions to date about the effects produced by vacuum, such as spontaneous emission, the Lamb shift,  the anomalous magnetic dipole moments of an electron, and etc.; all of them are under, at least in parts, direct influences of the vacuum. For the topics that has not been touched by the imaginary-time formalism, one of the important phenomena is the Casimir effect \cite{casimir}, which was theorized in 1948 as an attractive force in two conducting parallel plates. Without any background field, it was expected no force that could happen between two metal plates from the knowledge of the classical electromagnetism. In the picture of the quantum theory,  the empty  space is filled with all kinds of virtual particles, and the Casmir effect could be explained by the zero-point energy of the vacuum. The viable number of modes of the electromagnetic wave inside the two plates is much less than that in the outside, such that the exterior pressure is larger than the interior, therefore the net force is attractive. On the other hand, another effect of vacuum that has drawn much attentions from physicists is the van der Waals force \cite{vanderwaals}. In general, the term means  any attractive and repulsive forces between molecules other than those due to chemical bonds, or the electrostatic interaction.  They may include the forces between two permanent dipoles, a permanent and a corresponding  induced dipole, and two instantaneously induced dipoles. The last one is referred as the London dispersion force \cite{london}, and it is the one which accounts for the deviation of the ideal gas law and agrees with van der Waals' equation of state.  The origin of this minute force is from the polarizable particles due to the fluctuations of the zero-point energy of vacuum.  Experimentally, there have been many efforts made to test the predicted forces. The first attempt to test the Casimir force between two conducting plates was made by Sparnaay \cite{sparnaay} in 1958. Because of the large systematic errors and controllable electromagnetic forces, the result only shows qualitative agreement with the prediction. For decades, the research reaches some successful results in another geometry. For the setup constituted by a plane surface opposing a spherical surface \cite{mohideen}, the recent measurements have reached $1\%$ precision in the 0.5-3.0 $\mu$m range of distance. As for the test of two parallel conducting plates, in an experiment performed by Bressi {\it et al.} \cite{bressi} in 2002 the force coefficient was determined  at $15\%$ precision level. More recently, the vacuum photons radiated from the moving walls, the so-called dynamical Casimir effect, have been detected and confirmed \cite{paraoanu13}.  
\par 
Besides the above two vacuum effects, there is another one that has been under examination for decades and is of physicists' great interest. That is the Unruh effect \cite{unruh}, which was proposed by Fulling, Davies and Unruh respectively in the 1970s. It states that a uniformly accelerated observer will observe the black-body radiation. The theory is constructed by comparing the correlation function of a thermal bath of photons and that of photons measured by an observer with an acceleration $a$; an effective temperature measured by the accelerated detector in vacuum can be deduced: $T=\frac{\hbar a}{2\pi c k_B}$. Regardless of what experimental tests would tell us, it seems  counterintuitive  to imagine and describe the vacuum with thermodynamical characteristics.  In the recent researches, Akhmedov {\it et al.} \cite{akhmedov} found that a well known effect in synchrotron radiation  was actually a special case of the Unruh effect.  It is the so-called Sokolov–Ternov effect \cite{steffect}, which is an effect of the self-polarization of the accelerated electrons or positrons moving in a magnetic field. Once the anomalous magnetic dipole moments of the electrons are taken into consideration, the two effects can be found to be equivalent with the same relation for the temperature and the centripetal acceleration in a circular orbit. This confirms the existence of the Unruh effect, though more careful examinations are needed. In 1974, Hawking put forward a theory that black holes could evaporate near the surfaces of the event horizon, the so-called Hawking radiation \cite{hawking}. A similar relation of the black-body temperature of a black hole and the surface gravity, $\kappa$, close to the event horizon was concluded, similar to Unruh's, as $T=\frac{\hbar \kappa}{2\pi c k_B}$. Meanwhile, a thermodynamics of the black hole was established; one of the conclusions is that the surface area of the event horizon is identified as the entropy. Based from the above perspectives, it appears that in order to understand the gravity, which is the only theory that has not successfully quantized, its relationship with the vacuum as well as its thermodynamical features can not be separated.  An important role that could also be played by the vacuum is the dark energy. In another work \cite{huang13c}, the approaches of the DeWitt-Schwinger representation \cite{schwinger51,dewitt75} and the Casimir effect  to calculate the cosmological constant are adopted and extended. An equation of state of the vacuum, $p=w\rho$, with the ratio $w=-1$ can be obtained.  It has shown an integrated viewpoint  on the known vacuum effects and the applicability for reasonable results.
\par 
In the following section, the noted Casimir effect will be reviewed from the angle of the imaginary-time hamiltonian. The corresponding effects of the electromagnetic waves and fermions are discussed respectively. The implication to the van der Waals force will be presented in section 3. In section 4, the Unruh effect is recalculated from the imaginary-time two-point correlation function of the electromagnetic wave that is developed in ref. \cite{huang13a}. In section 5, the relation between the Hawking radiation and the thermal vacuum will be presented. A short conclusion is attached in the end. In principle, the natural units are adopted through the paper by setting $c=1$, $\hbar=1$, and etc., and they at times will be restored in the expressions without causing any ambiguity.

\section{Casimir effect}

\subsection{Electromagnatic Casimir effect}
\label{casimir}
In this section, the imaginary-time hamiltonian of virtual photons \cite{huang13a} is used to calculate energy between two parallel plates, the so called Casimir effect. It is known from the current knowledge of electromagnetism with the choice of gauge parameter $\zeta=1$ that the real-time hamiltonian is
\begin{eqnarray*}
\mathcal{H}_0(t,\vec{\bf x})=
\frac{1}{2}\sum_{k=1}^3\left[(\dot{A^k})^2+(\nabla A^k)^2\right]-
\frac{1}{2}\left[(\dot{A^0})^2+(\nabla A^0)^2\right],
\end{eqnarray*}
where the index $k$ only runs over spatial components from 1 to 3 and $\dot{A}$ means the time derivative of $A$, namely $\frac{d A}{dt}$. The imaginary-time hamiltonian is obtained from the above by transferring the time variable $t\rightarrow -i\tau$ and the integration $\int^\infty_{-\infty}dt$ is replaced by $\int^\beta_0d\tau$, where $\beta=\frac{1}{k_B T}$. The quantization of the photon fields can be fulfilled by the expansion of the Matsubara  frequency \cite{matsubara} for the imaginary-time. The hamiltonian density of the frequency $\omega_n$ and 3-momentum ${\bf q}=(q_x,q_y,q_z)$ is
\begin{eqnarray}
\mathcal{H}_0(\omega_n,\vec{\bf q})
&=&\frac{1}{4|{\bf q}|}(\omega_n^2-|{\bf q}|^2)\sum_\lambda (-g_{\lambda\lambda})
\left(a^\lambda_{\omega_n,{\bf q}} a^{\lambda\dagger}_{\omega_n,{\bf q}} + a^{\lambda\dagger}_{\omega_n,{\bf q}} a^\lambda_{\omega_n,{\bf q}}\right),\label{h0}
\end{eqnarray}
where $\lambda$ is the polarization index and $a^{\dagger}_{\omega_n,{\bf q}} $ and $ a^{\lambda\dagger}_{\omega_n,{\bf q}} $ are annihilation and creation operators of the photon field. The matrix $g_{\lambda\lambda'}={\rm diag}(1,-1,-1,-1)$, which is not a tensor, simply expresses the sign of each polarization state. The average energy of the system is obtained from 
\begin{eqnarray}
\langle \mathcal{H}_0\rangle &=&-\frac{\partial }{\partial \beta}\ln {\rm Tr} \,(e^{-\beta \mathcal{ H}_0})=\frac{{\rm Tr} \,( \mathcal{ H}_0e^{-\beta \mathcal{ H}_0})}{{\rm Tr} \,(e^{-\beta \mathcal{ H}_0})},\label{aveh0}
\end{eqnarray}
so that the average energy for the respective frequency and 3-momentum from eq. (\ref{h0}) is
\begin{eqnarray*}
\langle \mathcal{H}_0(\omega_n,\vec{\bf q})\rangle &=&-\frac{(i\omega_n)^2+|{\bf q}|^2}{(i\omega_n)^2-|{\bf q}|^2},
\end{eqnarray*}
where $\omega_n=\frac{2\pi n}{\beta}$ and $n$ is an integer for bosons. Then sum over all of the Matsubara frequencies and have the average energy density in 3-momentum phase space
\begin{eqnarray}
\langle \mathcal{H}_0(\vec{\bf q})\rangle &\equiv&\frac{1}{\beta}\sum_n\langle \mathcal{H}_0(\omega_n,\vec{\bf q})\rangle=-\frac{1}{\beta}\sum_n\frac{(i\omega_n)^2+|{\bf q}|^2}{(i\omega_n)^2-|{\bf q}|^2}.\label{aveh0s}
\end{eqnarray}
In order to attain the the sum of the above formula, consider a Fourier sum
\begin{eqnarray*}
G_B(\tau)=\frac{1}{\beta}\sum_n g_B(i\omega_n)e^{-i\omega_n\tau}.
\end{eqnarray*}
In our case, $g_B(i\omega_n)=-\frac{(i\omega_n)^2+|{\bf q}|^2}{(i\omega_n)^2-|{\bf q}|^2}$ and can be rewritten as 
 $g_B(i\omega_n)=-1-\frac{2|{\bf q}|^2}{(i\omega_n)^2-|{\bf q}|^2}$. The term of the minus unity contributes zero to $G_B$, since $\sum_n e^{-i\omega_n \tau}=0$. We may obtain 
\begin{eqnarray}
G_B(\tau)=|{\bf q}|\left({e^{-|{\bf q}|\tau}}{}+2\cosh (|{\bf q}|\tau) n_B(|{\bf q}|)\right),\label{Gb}
\end{eqnarray}
so that, as $\tau$ becomes infinitesimal and in the limit of zero temperature, the expression in eq. (\ref{aveh0s}) is summed as
\begin{eqnarray}
\langle  \mathcal{H}_0(\vec{\bf q})\rangle=\lim_{\tau\rightarrow 0^+}G_B(\tau) 
=\lim_{\tau\rightarrow 0^+}|{\bf q}|\left({e^{-|{\bf q}|\tau}}{}+2\cosh (|{\bf q}|\tau) n_B(|{\bf q}|)\right).\label{aveh0w}
\end{eqnarray}
It is important to keep the variable $\tau$ as nonzero, since it gives a limiting property to the sum: $\lim_{|{\bf q}|\rightarrow \infty} G_B(\tau)=0 $. And the reason will soon be clear in the following calculations. At this point, we may look back  on how the Casimir effect was calculated in the past. The conventional calculation of the Casimir effect \cite{casimir} starts from the zero-point energy in a cubical cavity of a length $L$:
\begin{eqnarray}
\sum_{l,m,n}(2)\frac{\hbar}{2} \omega_{lmn},\hspace{.5cm}{\rm where}\hspace{.5cm}\omega_{lmn}=\pi c \left[\frac{l^2}{L^2}+\frac{m^2}{L^2}+\frac{n^2}{L^2}\right]^{\frac{1}{2}}
\end{eqnarray}
and the factor 2 in the parenthesis arises from the two independent polarizations of modes. The above expression is identical to eq. (\ref{aveh0w})  in practical calculation, except the cutoff function $\left({e^{-|{\bf q}|\tau}}{}+2\cosh (|{\bf q}|\tau) n_B(|{\bf q}|)\right)$. For the physical situation of interest: two parallel plates are placed with a distance $d$, the summations of the allowed frequencies in two of the three dimensions become integrals for $l$ and $m$:  $\sum_{lmn}\rightarrow \sum_n (L/\pi)^2\int dq_x dq_y$. The energy of the electromagnetic waves in the cavity is 
\begin{eqnarray*}
E(d)=\frac{L^2}{\pi^2}(\hbar c)\sum_n\int^\infty_0 dq_x \int^\infty_0 dq_y\left(q^2_x+q^2_y+\frac{n^2 \pi^2}{d^2}\right)^{1/2},
\end{eqnarray*}
which apparently is an infinite quantity. If the two plates are placed infinitely far enough, $d\rightarrow \infty$, the sum over $n$ is also replaced by an integral. The potential energy in the two-plate system with a distance $d$ is the difference of the zero-point energies of the two situations: $U(d)=E(d)-E(\infty)$. The calculation of the potential function is proceeded by changing the two variables $q_x$ and $q_y$ to polar coordinates $r$ and $\theta$, so the potential becomes
\begin{eqnarray}
\hspace{-1cm}U(d)=\frac{L^2\hbar c}{\pi^2}\left(\frac{\pi}{2}\right)\left[\sum_n\int^\infty_0 dr \, r\left(r^2+\frac{n^2 \pi^2}{d^2}\right)^{1/2}-\left(\frac{d}{\pi}\right)\int^\infty_0 dq_z \int^\infty_0 dr\, r\left(r^2+q_z^2\right)^{1/2}\right].\label{aveh0si}
\end{eqnarray}
In the practical integration, a cutoff function, $f(|{\bf q}|)$, is intentionally added  into the above two integrals respectively for a convenience reason of the calculations. The cutoff function is required to have the properties: $f(|{\bf q}|)=1$ for $|{\bf q}|\ll  |{\bf q}_B|$ and $f(|{\bf q}|)=0$ for $|{\bf q}|\gg  |{\bf q}_B|$.  It is supposed that $|{\bf q}_B|\approx1/a_0$, where $a_0$ is the Bohr radius, due to the reason that the perfect conducting walls break down at small wavelengths. In fact, from the language of the imaginary-time hamiltonian, a cutoff function that already satisfies the two requirements is automatically provided as $f(|{\bf q}|)=\left({e^{-|{\bf q}|\tau}}{}+2\cosh (|{\bf q}|\tau) n_B(|{\bf q}|)\right)$ in eq. (\ref{aveh0w}) and $\tau$ can be chosen to be $a_0$, while $\beta\gg 1$ for a nearly zero temperature, the second term of $f(|{\bf q}|)$ can be neglected. However, the appearance of this term will show its importance as we calculate its energy density constant  and pressure through the same manner. In the other limit of $d$ going to infinity, this term will become dominant and lead to the equation of state with a pressure-to-density ratio $w=-1$, this is what we expect for the dark energy \cite{huang13c}.  In order to derive the difference of the sum and integral, the Euler-Maclaurin summation formula \cite{eulermaclaurin}, which reads
\begin{eqnarray}
\sum^{\infty}_{n=1}F(n)-\int^\infty_0 dx F(x)=-\frac{1}{2}F(0)-\frac{1}{12}F'(0)+\frac{1}{720}F'''(0)+\dots,
\end{eqnarray}
 is used to compute the terms in the square bracket of eq.  (\ref{aveh0si}). The two required properties stated above for the cutoff function are necessary for applying this formula. Following the same derivation as in ref. \cite{casimir}, the potential energy can be obtained, $U(d)=-\frac{\pi^2\hbar c}{720 d^3}L^2$, and the Casimir force per unit area is $\frac{F(d)}{L^2}=-\frac{\pi^2\hbar c}{240 d^4}$. The value of the imaginary-time formalism on the Casimir effect is that it not only gives consistent results but provides a cutoff function. To find out more  physical meanings of this cutoff function $f(|{\bf q}|)$, we may re-examine another approach of the Casimir effect by Lifshitz \cite{lifshitz},  which considers the energy of the vacuum field in a dielectric slab between two parallel plates. Recall the average energy of induced $N$ dipole ${\bf d}$ per unit volume is $\langle E\rangle=-\frac{1}{2}\int d^3 r\langle {\bf P}\cdot {\bf E}\rangle $, where ${\bf P}\,(=N {\bf d})$ is the polarization. Taking into account the normal ordering of the field operators, the expectation value of the energy stored in the dielectric slab becomes
\begin{eqnarray}
\langle E\rangle=-\frac{1}{2}\int d^3 r\langle {\bf P}\cdot {\bf E}^{(-)}+{\bf P}\cdot {\bf E}^{(+)}\rangle,\label{aveEP}
\end{eqnarray}
where ${\bf E}^{\pm}$ are the positive- and negative-frequency parts of the source-free electromagnetic fields. The polarization can be written as 
\begin{eqnarray*}
{\bf P}({\bf r},t)=N({\bf r})\int d \omega \alpha(\omega)\left(  {\bf F}^{(+)}({\bf r},\omega)e^{-i\omega t}+ {\bf F}^{(-)}({\bf r},\omega)e^{i\omega t}\right),
\end{eqnarray*}
where $\alpha(\omega)$ is the polarizability and ${\bf E}^{(\pm)}({\bf r},t)=\int d\omega {\bf F}^{(\pm)}({\bf r},\omega) e^{\mp i\omega t}$. Substitute the above into eq. (\ref{aveEP}), it becomes 
\begin{eqnarray}
\langle E\rangle=-\frac{1}{2}\int d^3 r N({\bf r})\int d\omega \alpha(\omega ) \langle {\bf F}^{(+)}\cdot {\bf F}^{(-)}+{\bf F}^{(-)}\cdot {\bf F}^{(+)}\rangle.\label{aveEPa}
\end{eqnarray}
We know that $\langle {\bf F}^{(+)}\cdot {\bf F}^{(-)}\rangle$ is proportional to $\langle a_\omega (0) a^\dagger_\omega (0)\rangle $ and  $\langle{\bf F}^{(-)}\cdot {\bf F}^{(+)}\rangle$ is proportional to $\langle a^\dagger_\omega (0) a_\omega (0)\rangle $ similarly; those are, in a finite temperature, $1+n_B(\omega)$ and $n_B(\omega)$. In the limiting case of setting the dielectric constant to be unity, the expression in eq. (\ref{aveEPa}) reduce to the average energy of the vacuum field and the result for the Casimir effect can be retrieved. 
\subsection{Fermion Casimir effect}
\label{fcasimir}
Now we may turn our attention to fermions, consider the Dirac field in two parallel plates at $z=0$ and $z=d$. As in the case of the electromagnetic waves, the boundary conditions of the fermion currents, $j^\mu(x)=(j^0,{\bf j})$, are necessary. On the surfaces of the two plates, $\hat{\bf n}\cdot {\bf j}(x)=0$, where $\hat{\bf n}$ is the unit vector normal to the surface, such that no particle could penetrate through the walls. The condition can be written in a covariant form, $n_\mu j^\mu=0$, where $n^\mu=(0,\hat{\bf n})$. A positive-energy Dirac wave function takes the form \cite{chodos}
\begin{eqnarray}
\psi(x,t)=e^{-i\xi_{\bf p}t}e^{ip_x x+ip_yy}\left(e^{ip_zz}+i\gamma^3 e^{-ip_zz}\right)u({\bf p}),
\end{eqnarray}
where $u({\bf p})$ is the positive-energy spinor. The boundary condition requires that $p_z=\frac{n\pi}{2d}$, where $n$ is an odd number. Therefore, the energy of each Dirac particle is
\begin{eqnarray*}
\xi_{\bf p}=|{\bf p}|=\sqrt{p^2_x+p^2_y+\frac{n^2\pi^2}{4d^2}}.
\end{eqnarray*}
The hamiltonian of fermions in the imaginary-time formalism from ref. \cite{huang13a} is 
\begin{eqnarray}
\mathcal{H}_{\rm D}(\omega_n, \vec{\bf p})
&=&\sum_{s}\left.\xi_{\bf p}{a^s}^\dagger_{\omega_n,{\bf{p}}}a^s_{\omega_n,{\bf{p}}}\right.
+\left.\sum_{s}\xi_{\bf p}b^s_{\omega_n,\bf p}{b^s}^\dagger_{\omega_n,\bf p}\right. .\label{Lagrangian2}
\end{eqnarray}
Similarly, the average energy can be also obtain from eq. (\ref{aveh0}) with $\mathcal{H}_{0}$ replaced by $\mathcal{H}_{\rm D}$, so that when summing over all of the fermionic Matsubara frequencies
\begin{eqnarray}
\langle\mathcal{H}_{\rm D}(\vec{\bf p})\rangle
=\frac{1}{\beta}\sum_n\langle\mathcal{H}_{\rm D}(\omega_n, \vec{\bf p})\rangle
&=&\frac{1}{\beta}\sum_n\frac{2\xi_{\bf p}}{i\omega_n-\xi_{\bf p}}-\frac{1}{\beta}\sum_n\frac{2\xi_{\bf p}}{i\omega_n+\xi_{\bf p}},
\end{eqnarray}
where the factor 2 is from the up- and down-spins, an exponential factor, $e^{-i\omega_n\tau}$, with $\tau\rightarrow 0^+$ is introduced. Then we may obtain 
\begin{eqnarray}
\langle\mathcal{H}_{\rm D}(\vec{\bf p})\rangle
=\lim_{\tau\rightarrow 0^+}\frac{1}{\beta}\sum_{n={\rm odd}}\frac{4\xi^2_{\bf p}e^{-i\omega_n\tau}}{(i\omega_n)^2-\xi^2_{\bf p}}
=\lim_{\tau\rightarrow 0^+}-2\xi_{\bf p}\left(e^{-\xi_{\bf p}\tau}-2\cosh(\xi_{\bf p}\tau)n_F(\xi_{\bf p})\right).\label{hamD}
\end{eqnarray}
The above formula is the exact expression of the zero-point energy of Dirac fields in the field theory except the terms in the parenthesis.  If the temperature of the vacuum is close to zero, the term with the density function can be neglected. As in the bosonic case, this second  term in the parenthesis of eq. (\ref{hamD}) cannot be thrown away for a extremely large $d$, and it also contributes some to the cosmological constant, though not as much as in photon's case \cite{huang13c}.  Integrate and sum over the 3-momenta, the average energy, $E(d)$ , is obtained as
 \begin{eqnarray}
E(d)=\lim_{\tau\rightarrow 0^+}-2\sum_{n={\rm odd}}\int\frac{d^2{\bf p}_{\bot}}{(2\pi)^2}\sqrt{{\bf p}^2_\bot+\frac{n^2\pi^2}{4d^2}} e^{-\sqrt{{\bf p}^2_\bot+\frac{n^2\pi^2}{4d^2}}\tau},\label{aveED}
\end{eqnarray}
where ${\bf p}_\bot=(p_x,p_y)$. The reason that the exponential factor is still kept in the above is that it is given to regularize the divergence in the zero-point energy. With the variable change $x=\sqrt{{\bf p}^2_\bot+\frac{n^2\pi^2}{4d^2}}$, $E(d)$ becomes 
 \begin{eqnarray*}
E(d)=\lim_{\tau\rightarrow 0^+}-\frac{1}{\pi}\sum_{n={\rm odd}} \int^\infty_{\frac{n\pi}{2d}}x^2 dxe^{-x\tau}=\lim_{\tau\rightarrow 0^+}-\frac{1}{\pi}\frac{\partial^2 }{\partial \tau^2}\sum_{n={\rm odd}} \int^\infty_{\frac{n\pi}{2d}} dxe^{-x\tau}\\
=\lim_{\tau\rightarrow 0^+}-\frac{1}{\pi}\frac{\partial^2 }{\partial \tau^2}\frac{1}{\tau}\sum_{n={\rm odd}} e^{-\frac{n\pi}{2d}\tau},\hspace{.45cm}
\end{eqnarray*}
where outcome of  the summation is $\sum_{n{\rm }} e^{-\frac{n\pi}{2d}\tau} =\left(2\sinh\frac{\pi\tau}{2d}\right)^{-1}=\frac{d}{\pi\tau}-\frac{1}{12}\frac{\pi\tau}{2d}+\frac{7}{720}\frac{\pi^3\tau^3}{8d^3}+O(\tau^4)$ for odd numbers of $n$. In the end, we may ignore the divergent term as the limit of $\tau\rightarrow 0^+$ is taken. This is legitimate since the potential is defined as $U(d)\equiv E(d)-E(\infty)$, the divergence will be removed by taking the difference. In fact, the divergence only happens due to some flaws in the above derivation, and it will no longer exist if a good value of $\tau$ is chosen. More detail and the related physical  reason can be found in ref. \cite{huang13c}.   The finite term left for the potential and the induced Casimir force for the Dirac fields are 
 \begin{eqnarray*}
U(d)=-\frac{7\pi^2}{2880d^3},\,\,{\rm and}\hspace{.5cm}F(d)=-\frac{7\pi^2}{960d^4}.
\end{eqnarray*}
Most of the derivations starting from eq. (\ref{aveED}) are the same as those in ref. \cite{casimir}. The difference is that the factor, $e^{-\xi_{\bf p}\tau}$, is added in their calculations 
on purpose. In short, as we may observe from the above computations, the hamiltonian density of the imaginary-time formalism  generate  consistent results with the precedent works for the fermionic Casimir effect, just like  the  case of electromagnetic waves. Moreover, from this imaginary-time approach both kinds of vacuum field are automatically provided with cut-off functions $e^{-|{\bf q}|\tau}$ and $e^{-\xi_{\bf p}\tau}$ for the physical observables.

\section{Van der Waals force }
\label{vdW}
In 1873, van der Waals \cite{vanderwaals} proposed the equation of state modified from the ideal gas law, for one mole of a gas at a temperature $T$
\begin{eqnarray*}
\left(P+\frac{a}{V^2}\right)(V-b)=R\,T,
\end{eqnarray*}
where $P$, $V$ and $R$ are the pressure, volume and the gas constant. The parameter $a$ and $b$ are called the van der Waals constants, and are determined by fitting to the experimental data. The constant $b$ was interpreted by van der Waals as the volume occupied by two atoms; the constant $a$ was related  to an attractive force between two atoms. Since then, many attempts were made to suggest that the correction to the ideal gas law was because of the interaction potential, such as that of two molecules with permanent dipole moments \cite{keesom} or that of one permanent quadruple and one induced dipole moments \cite{debye}. However, it was soon recognized that gases of non-polar molecules have nonzero values of the van der Waals constants $a$, and neither force is capable of explaining the van der Waals equation of state. In 1930, London \cite{london} derived the interaction potential of two atoms or molecules between the ground and the first excited states from the perturbation theory of quantum mechanics. He showed that a dipole moment could be induced in each molecule even if neither of two molecules has a permanent moment. 
Besides, London's force, unlike Keesom's for two permanent moments, is temperature independent. Consider that the total electric field acting on the  atom $A$ is divided into two components  
\begin{eqnarray*}
{\bf E}({\bf x},t)&=&{\bf E}_B({\bf x},t)+{\bf E}_0({\bf x},t),
\end{eqnarray*}
where ${\bf E}_B$ is the field exerted by the second atom $B$ and ${\bf E}_0$ is the zero-point field. The electric field is obtained by using ${\bf E}=-i\frac{\partial }{\partial \tau}{\bf A}-\nabla A^0$, where the imaginary-time is applied. The quantization of the electric field  therefore can be made through the electromagnetic vector potential $A^\mu=(A^0,{\bf A})$. The zero-point field is expanded according to the imaginary-time formalism by the Matsubara frequency and 3-momentum
\begin{eqnarray*}
\hspace{-.8cm}{\bf E}^i_0({\bf x},\tau)&\equiv&\frac{1}{\beta}\sum_n\int\frac{d^3{\bf k}}{(2\pi)^3}\sum_{\lambda'} \mathcal{ E}^i_{\lambda',0}(\omega_n,{\bf k};\tau,{\bf x})\\
&=&\frac{1}{\beta}\sum_n\int\frac{d^3{\bf k}}{(2\pi)^3}\sum_{\lambda'}\frac{ \omega_n}{\sqrt{2|{\bf k}|}}\left(- a^{\lambda'}_{\omega_n,{\bf k}}e^{-i\omega_n\tau+i{\bf k\cdot x}} \epsilon^i_{\lambda'}({\bf k})+a^{\lambda'\dagger}_{\omega_n,{\bf k}}e^{i\omega_n\tau-i{\bf k\cdot x}} \epsilon^i_{\lambda'}({\bf k})\right),
\end{eqnarray*}
where $\lambda'$ are two transverse components of the polarization states. And the energy of atom A in the electric field is 
\begin{eqnarray*}
W_A=-\frac{1}{2\beta}\sum_{n,\lambda'}\int \frac{d^3{\bf k}}{(2\pi)^3}\alpha_A({\bf k})|{\mathcal E}_{\lambda',0}(\omega_n,{\bf k};\tau,{\bf x})|^2,
\end{eqnarray*}
where $\alpha_A({\bf k})$ is the polarizability of atom A at ${\bf x}$ and $\mathcal{E}_{\lambda',0}=(\mathcal{E}^i_{\lambda',0},\mathcal{E}^2_{\lambda',0},\mathcal{E}^3_{\lambda',0})$ is a 3-vector. Extract the part of $W_A$ for the interaction between the two atoms, and its average is
\begin{eqnarray*}
\hspace{-.8cm}\langle W_{AB}\rangle=-\frac{1}{2\beta}\sum_{n,\lambda'}\int \frac{d^3{\bf k}}{(2\pi)^3}\alpha_A({\bf k})\left(\langle {\mathcal E}^{(+)}_{\lambda',0}\cdot {\mathcal E}_{\lambda',B} \rangle+\langle{\mathcal E}_{\lambda',B} \cdot {\mathcal E}^{(-)}_{\lambda',0}  \rangle +\langle {\mathcal E}^{(-)}_{\lambda',0}\cdot {\mathcal E}_{\lambda',B} \rangle+\langle{\mathcal E}_{\lambda',B} \cdot {\mathcal E}^{(+)}_{\lambda',0}  \rangle\right).
\end{eqnarray*}
Notice that $\mathcal{E}_{\lambda',B}$  is the electric field of the real photons and should not be expressed  in the imaginary -time formalism. However, the source of $\mathcal{E}_{\lambda',B}$ is the dipole moment of the atom $B$, ${\bf p}_B$, which is induced by the electric field of the zero-point. The analytic continuation to the imaginary-time has to be applied at some point, and this is the reason why $\mathcal{E}_{\lambda',B}$ can be directly inserted in the above. From the classical electromagnetism \cite{jackson}, the electric field generated by atom $B$'s dipole moment ${\bf p}_B=\hat{\mu}_B p_B(t)$, where $\hat{\mu}_B$ is the unit vector of ${\bf p}_B$, is 
\begin{eqnarray}
\hspace{-.8cm}{\bf E}_B({\bf x},t)=-\left[\hat{\mu}_B-(\hat{\mu}_B\cdot {\bf s}){\bf s}\right]\frac{1}{r}\ddot{\bf p}_B(t-r)+
\left[3(\hat{\mu}_B\cdot {\bf s}){\bf s}-\hat{\mu}_B\right]\left[\frac{1}{r^3}{\bf p}_B(t-r)+\frac{1}{r^2}\dot{\bf p}_B(t-r)\right],\label{eb}
\end{eqnarray}
where ${\bf s}$ is the unit vector from the atom $B$ to the atom $a$ and $r$ is the distance between the two. The dots one the top of ${\bf p}_B$ mean the time derivatives. As what was emphasized earlier,  the induced dipole moment of atom $B$ due to the zero-point field can be expressed for the real-time as 
\begin{eqnarray}
{\bf p}_B({\bf y},t)=\int \frac{d^3{\bf k}}{(2\pi)^3}\sum_{\lambda'}\alpha_B({\bf k})\left[\mathcal{E}^{(+)}_{\lambda',0}({\bf k},{\bf y},t)+\mathcal{E}^{(-)}_{\lambda',0}({\bf k},{\bf y},t)\right].\label{pb}
\end{eqnarray}
After substituting the above into eq. (\ref{eb}), we may obtain the electric field generated by the atom $B$ for the positive and negative frequencies respectively as below
\begin{eqnarray*}
&&\hspace{-.8cm}\mathcal{E}^{(+)}_{\lambda',B}({\bf k}, {\bf x}, t)=\alpha_B({\bf k})
\hat{\mathcal{R}}\cdot \mathcal{E}^{(+)}_{\lambda',0}({\bf k}; t, {\bf y}),\hspace{1.cm}
\mathcal{E}^{(-)}_{\lambda',B}({\bf k}, {\bf x}, t)=\alpha_B({\bf k})
\hat{\mathcal{R}}\cdot \mathcal{E}^{(-)}_{\lambda',0}({\bf k}; t, {\bf y}),
\\
&&\hspace{-0.7cm}{\rm where}\\
&&\hspace{-1.5cm}\hat{\mathcal{R}}_\pm=|{\bf k}|^3e^{\pm i|{\bf k}| r}\left\{\frac{\left.{\vec \epsilon}_{\lambda'}({\bf k})-({\vec \epsilon}_{\lambda'}({\bf k})\cdot {\bf s}){\bf s}\right.}{|{\bf k}|r}+\left[3({\vec \epsilon}_{\lambda'}({\bf k})\cdot {\bf s}){\bf s}-{\vec \epsilon}_{\lambda'}({\bf k})\right]\left(\frac{1}{|{\bf k}|^3r^3 }\mp\frac{i}{|{\bf k}|^2r^2 }\right)\right\}\otimes {\vec\epsilon}_{\lambda'}({\bf k}).
\end{eqnarray*}
The operators $\hat{\mathcal{R}}_{\pm}$ are  matrices of $3\times 3$ spatial indices. Emphasize again that the above expressions of the electric field from the atom $B$ are derived from eq. (\ref{eb}) and (\ref{pb}) for the real-time, however we may regard the terms in front of the zero-point fields as linear operators on $\mathcal{E}^{(\pm)}_{\lambda',0}({\bf k};t, {\bf y})$, which are proportional to the plane-wave eigenfunctions $\propto e^{\mp i|{\bf k}|t\pm i{\bf k\cdot x}}$. In the imaginary-time formalism,  the fields are expanded by the Matsubara frequency and the usual 3-momentum $\propto e^{\mp i\omega_n\tau\pm i{\bf k\cdot x}}$. As  deriving the propagator of the photons in ref. \cite{huang13a}, the factor $e^{\pm  |{\bf k}|\tau }$ will appear after all of the mastsubara frequencies  are summed, and they become $e^{\pm i |{\bf k}| t }$ after the analytic continuation is applied. Thus we may use the imaginary-time eigenfunctions, which are now denoted as $\mathcal{E}^{(\pm)}_{\lambda',0}(\omega_n,{\bf k};\tau, {\bf y})$, in obtaining $\mathcal{E}^{(\pm)}_{\lambda, B}$. The interaction energy becomes
\begin{eqnarray*}
&&\hspace{-1.5cm}\langle W_{AB}\rangle=-\frac{1}{2\beta}\sum_{n,\lambda'}\int \frac{d^3{\bf k}}{(2\pi)^3}\alpha_A({\bf k}) \alpha_B({\bf k})\left\{  {\vec \epsilon}_{\lambda'}({\bf k})\cdot \left( \hat{\mathcal{R}}_++ \hat{\mathcal{R}}_-\right)\cdot {\vec \epsilon}_{\lambda'}({\bf k})
\left( \langle \mathcal{E}^{(+)}_{\lambda',0} \cdot  \mathcal{E}^{(-)}_{\lambda',0} \rangle+ \langle \mathcal{E}^{(-)}_{\lambda',0} \cdot  \mathcal{E}^{(+)}_{\lambda',0} \rangle\right)\right\}.\hspace{.5cm}
\end{eqnarray*}
After summing over the Matsubara frequencies, we may obtain   
\begin{eqnarray*}
&&\hspace{-1.4cm}\frac{1}{\beta}\sum_n\langle \mathcal{E}^{(+)}_{\lambda',0} \cdot  \mathcal{E}^{(-)}_{\lambda',0}\rangle=\frac{1}{\beta}\sum_n\frac{(i\omega_n)^2}{2|{\bf k}|}\langle a^{\lambda'}_{\omega_n,{\bf k}}a^{\lambda',\dagger}_{\omega_n,{\bf k}}  \rangle=-\frac{1}{\beta}\sum_n\frac{(i\omega_n)^2}{(i\omega_n)^2-{\bf k}^2}\nonumber\\
&&\hspace{5.5cm}=\lim_{\tau\rightarrow 0^+}\frac{|{\bf k}|}{2}\left({e^{-|{\bf k}|\tau}}{}+2\cosh (|{\bf k}|\tau) n_B(|{\bf k}|)\right),
\end{eqnarray*}
where $\tau$, like in eq. (\ref{Gb}) and (\ref{hamD}) , is introduced to help the summation. As we will realized soon later, it also provides to regularize the divergence just like in the case of the Casimir effect. The other expectation value $\langle \mathcal{E}^{(-)}_{\lambda',0} \cdot  \mathcal{E}^{(+)}_{\lambda',0} \rangle$ can be derived in the same way, and the result is identical with the previous one. In the limit of zero temperature, the bosonic density function $n_B(|{\bf k }|)$ can be ignored. The integral is proceeded with the unit vector $\hat{\bf s}=(0,0,1)$ pointing at $z$-direction, 
\begin{eqnarray*}
&&\hspace{-1.5cm}\langle W_{AB}\rangle=\lim_{\tau\rightarrow 0^+}-\frac{1}{\pi}\int^\infty_0 d|{\bf k}| {\bf k}^6\alpha_A({\bf k})\alpha_B({\bf k})G(|{\bf k}|r)e^{-|{\bf k}|\tau},\hspace{.4cm}\nonumber\\
&&{\rm where}\,\,\,\,G(x)\equiv \frac{\sin 2 x}{x^2}+\frac{2\cos 2 x}{x^3}-\frac{5\sin 2 x}{x^4}-\frac{6\cos 2 x}{x^5}+\frac{3\sin 2 x}{x^6}.
\end{eqnarray*}
The above result is the same as  that in the second reference of \cite{casimir} except the exponential factor of regularization. The cut-off function was added in the above formula by Casimir and Polder \cite{casimirpolder} in 1948 because of retardation. For the larger distance between atoms, the retardation becomes important. By assuming the polarizabilities $\alpha_A$ and $\alpha_B$ are constants, we may calculate the interacting potential $V(r)=\langle W_{AB}\rangle$,
\begin{eqnarray*}
V(r)&=&\lim_{\tau\rightarrow 0^+}-\frac{\alpha_A\alpha_B}{\pi}\int^\infty_0 d|{\bf k}| {\bf k}^6G(|{\bf k}|r) e^{-|{\bf k}|\tau}=-\frac{23}{4\pi r^7}\alpha_A\alpha_B.
\end{eqnarray*}
Since the van der Waals potential is required to fall off faster than $1/r^6$ in order for theory and experiment to be consistent \cite{verwey}, this is the only one that is qualified among the those attempts to explain the van der Waals force.  

\section{Unruh effect }
\label{unruh}
The Unruh effect \cite{unruh} states that a detector with a uniform acceleration $a$ in the vacuum responds as if it were at rest in a thermal bath at a temperature $T=\frac{\hbar a}{2\pi c k_B}$. In principle,  the equivalence principle \cite{einstein} in general relativity tells that physics laws in a gravitation field can not be differentiated from those in an accelerated frame. It implies that the acceleration $a$ can be identified as a gravitation field $g$. As a result,
it seems that a gravitation field $g$ also corresponds to a secret temperature $T$ with the acceleration $a$ replaced by a gravity $g$ in the relation.  In fact, it is also found in theory that a black hole could radiate from its event horizon like a black-body of a temperature,  $T=\frac{\hbar \kappa}{2\pi c k_B}$, where $\kappa$ is the surface gravity on the event horizon.  The larger a gravity is, the higher the temperature will be. And this is the same expression of the temperature after the equivalent principle is applied in the Unruh effect. It is the so-called Hawking radiation \cite{hawking}. However, the predicted temperature,  $\approx 10^{-6}\left(\frac{M_\odot}{M}\right)K$, is so small that any measurement is a difficult task. In the theory of the imaginary-time field theory, it  provides a ready explanation for the imaginary thermal bath that is implied by  those effects.
Therefore  the similarities between the imaginary-time electromagnetic theory and the Unruh effect are worth a deeper exploration. Following a similar way in discussing the Unruh effect, the retarded two-point correlation function of the electromagnetic waves from the imaginary-time formalism \cite{huang13a} at the same spacial point are 
\begin{eqnarray}
&&\hspace{-1cm}\left\langle A_\mu(\tau_x,{\bf x}) A_\nu(\tau_y,{\bf x})\right\rangle
=\frac{1}{\beta}\sum_{n}\int \frac{d^3 {\bf q}}{(2\pi)^3}\frac{-g_{\mu\nu}}{2|{\bf q}|}\left\{\frac{1}{i\omega_n+|{\bf q}|}-
\frac{1}{i\omega_n-|{\bf q}|} \right\}e^{-i\omega_n(\tau_x-\tau_y)},\nonumber\\
&&\hspace{2.25cm}=\int \frac{d^3 {\bf q}}{(2\pi)^3}\frac{-g_{\mu\nu}}{2|{\bf q}|}\left\{\frac{e^{-|{\bf q}|\Delta\tau}}{e^{\beta |{\bf q}|}-1}+
\left(1+\frac{1}{e^{\beta |{\bf q}|}-1} \right)e^{|{\bf q}|\Delta\tau}\right\},\nonumber\\
&&\hspace{2.25cm}= \frac{-g_{\mu\nu}}{4\pi^2}  \int^\infty_0{|\bf q|} { d|{\bf q}|}\left({e^{-|{\bf q}|\Delta\tau}}+
\frac{2\cosh(|{\bf q}|\Delta \tau)}{e^{\beta |{\bf q}|}-1} \right),\nonumber\\
&&\hspace{2.25cm}=\frac{-g_{\mu\nu}}{4\beta^2} \csc^2\left(\frac{\pi\Delta\tau}{\beta}\right )\xrightarrow{\Delta\tau\rightarrow i\Delta t} \,\,\frac{g_{\mu\nu}}{4\beta^2} {\rm csch}^2\left(\frac{\pi\Delta t}{\beta}\right ),
\label{PhotonProp1}
\end{eqnarray}
where $\Delta \tau=\tau_y-\tau_x$. In the last step, the analytic continuation of the imaginary-time, $\tau\rightarrow it$, is applied. In the zero-temperature limit between any two space-time points, consider the two-point correlation 
\begin{eqnarray}
&&\hspace{-2.2cm}\left\langle A_\mu(\tau_x,{\bf x}) A_\nu(\tau_y,{\bf y})\right\rangle
=\frac{1}{\beta}\sum_{n}\int \frac{d^3 {\bf q}}{(2\pi)^3}\frac{-g_{\mu\nu}}{2|{\bf q}|}\left\{\frac{1}{i\omega_n+|{\bf q}|}-
\frac{1}{i\omega_n-|{\bf q}|} \right\}e^{-i\omega_n(\tau_x-\tau_y)+i{\bf q }\cdot ({\bf x-y})},\nonumber\\
&&\hspace{1cm}=\int \frac{d^3 {\bf q}}{(2\pi)^3}\frac{-g_{\mu\nu}}{2|{\bf q}|}\left\{\frac{e^{-|{\bf q}|\Delta\tau+i{\bf q }\cdot ({\bf x-y})}}{e^{\beta |{\bf q}|}-1}+
\left(1+\frac{1}{e^{\beta |{\bf q}|}-1} \right)e^{|{\bf q}|\Delta\tau+i{\bf q }\cdot ({\bf x-y})}\right\},\nonumber\\
&&\hspace{0,5cm}\xrightarrow{\beta\rightarrow \infty}\int \frac{d^3 {\bf q}}{(2\pi)^3}\frac{-g_{\mu\nu}}{2|{\bf q}|}
e^{i|{\bf q}|\Delta t+i{\bf q }\cdot \Delta {\bf x}}=-\frac{g_{\mu\nu}}{8\pi^2}\int |{\bf q}|d|{\bf q}|d\cos\theta
e^{i|{\bf q}|\Delta t+i{\bf q }\cdot \Delta {\bf x}},\nonumber\\
&&\hspace{1cm}=-\frac{g_{\mu\nu}}{4\pi^2}\int |{\bf q}|d|{\bf q}|
e^{i|{\bf q}|\Delta t}\frac{\sin(|{\bf q}||\Delta {\bf x}|)}{|{\bf q}||\Delta{\bf x}|}=\frac{g_{\mu\nu}}{4\pi^2}\frac{1}{\Delta{t}^2-\Delta {\bf x}^2},
\label{PhotonProp2}
\end{eqnarray}
 where $\Delta {\bf x}={\bf y-x}$. After taking the limit of the zero temperature and the analytic continuation,  the last line is obtained. For a uniformly accelerating observer in two dimensional space-time, the trajectory can be derived by the Rindler's metric \cite{rindler}: ${\bf x}(t)=\frac{1}{a}\left( \cosh(a t)-1\right)$, where $a$ is the acceleration. Not to repeat the calculations that are known already, the resultant two-point correlation function of eq. (\ref{PhotonProp2}) is
\begin{eqnarray}
\hspace{-1cm}\left\langle A_\mu(t_x,{\bf x}) A_\nu(t_y,{\bf y})\right\rangle&=& \frac{g_{\mu\nu} }{16\pi^2}a^2{\rm csch}^2\left(\frac{ a\Delta t}{2}\right).\label{PhotonProp3}
\end{eqnarray}
Compare eq. (\ref{PhotonProp1}) with ({\ref{PhotonProp3}}), the Unruh effect is derived as $a=\frac{2\pi}{\beta}$. It  is interpreted as  that a uniformly accelerated detector sees a thermal bath in the move. 
As for the gravitational fields, the corresponding temperature can be viewed as the intrinsic temperature of the vacuum, since the gravity is well-defined and  is only determined by the position in the space-time.

\section{Hawking radiation}
\label{hawking}
The Hawking radiation \cite{hawking} is theorized for the presence of a black hole, which could be static, rotating or charged. Similar mathematical derivations could also be imitated for a system of an accelerated mirror \cite{birrel}. In principle, they are derived through matching the outgoing and ingoing rays of light during the collapse of the black hole. One important feature of the black-hole dynamics is that Wightman functions \cite{birrel} possess thermodynamical characteristics. It is well known that the thermal Green functions are periodic with a period $\beta\, (=\frac{1}{k_B T})$ with respective to the imaginary-time axis, the so-called KMS condition \cite{matsubara} in the statistical quantum  mechanics. 
The vacuum states are defined in different circumstances, for instance, the field quantizations are given separately inside the event horizon and outside in the distance. Namely, vacuum state $|0\rangle$ is defined for the operator $a_{\bf k}$, and $|\bar{0}\rangle$ for $\bar{a}_{\bf k}$. The transformation between the fields defined in distinct vacuums will inevitably introduce the conformal invariance to the action as the Tolman-Ehrehfest relation \cite{tolman30} is applied. Thus the physical scale is determined by the thermal property of the vacuum. More explanations can be found in \cite{huang13a,huang13c} and the references therein. With the Bogoliubov transformation \cite{matsubara,birrel}  between the modes of the ingoing and the outgoing wave functions,   the particle number per mode is obtained as
\begin{eqnarray*}
\frac{1}{e^{8\pi M \omega}-1},
\end{eqnarray*}
where $M$ is the mass of the black hole and $\omega$ is the frequency of the electromagnetic wave. The surface gravity of a black hole, $\kappa$, at the event horizon is $\frac{1}{4M}$, thus the Hawking's temperature can be acquired as $T=\frac{\kappa}{2\pi c k_B}$. It is derived from the Schwarzschild metric and Kruskal coordinates of a static black hole, and is often illustrates by the Penrose diagrams \cite{MTW}. Many details can be found in the references that have been mentioned earlier. As the vacuum is pictured as a finite-temperature continuum, and  photons that are generated from a black hole behave like the black body radiation.  A black hole provides a  unique environment near the event horizon for a half of the virtual particles to become genuinely real while the other half fall into the singularity. In the following, we may briefly describe its relation to the imaginary-time Green functions. Consider a scalar field that is expanded by two orthonormal sets of modes, $u_{\bf k}(x)$ and $\bar{u}_{\bf k}(x)$
\begin{eqnarray*}
\phi(x)=\int \frac{d^n{\bf k}}{(2\pi)^3} a_{\bf k}u_{\bf k}(x)+a^*_{\bf k}u^*_{\bf k}(x),\hspace{.1cm}{\rm and}\hspace{.3cm}
\phi(x)=\int \frac{d^n{\bf k}}{(2\pi)^3} \bar{a}_{\bf k}\bar{u}_{\bf k}(x)+\bar{a}^*_{\bf k}\bar{u}^*_{\bf k}(x),
\end{eqnarray*}
where $n$ is the number of the spatial dimensions. For a spherical black hole, the modes of the wave functions are the products of the radial and the polar angle parts, so the discussion can be simplified to 2 dimensions of space-time (t,r) for radial photons. The mode $\bar{u}_{\bf k}(x)$ is for the outgoing wave functions and $u_{\bf k}(x)$ for the ingoing wave functions, and $x=(t,{\bf x})$. In Hawking's theory \cite{hawking}, the plane waves of the outgoing modes, $\bar{u}_{\bf k}(x)=\frac{1}{\sqrt{2\omega}} {e^{-i\omega t+i{\bf k\cdot x}}}$, are related to the ingoing modes by a Bogoliubov transformation:
\begin{eqnarray}
{u}_{\bf k}(x)=\sum_{\omega'}\alpha_{{\omega, \omega'}}\,\bar{u}_{\bf k'}(x)+\beta_{{\omega, \omega'}}\,\bar{u}^*_{\bf k'}(x),\label{bogo}
\end{eqnarray}
where $\omega=|{\bf k}|$. The coefficients $\alpha_{{\omega, \omega'}}$ and $\beta_{{\omega, \omega'}}$ are the inner products of $({u}_{\bf k}(x),\bar{u}_{\bf k'}(x))$ and $({u}_{\bf k}(x),\bar{u}^*_{\bf k'}(x))$.  And they become
$\alpha_{{\omega, \omega'}}=\alpha_\omega\delta_{\omega\omega'}$ and $\beta_{{\omega, \omega'}}=\beta_\omega\delta_{\omega,-\omega'}$, since ${u}_{\bf k}(x)$ and $\bar{u}_{\bf k}(x)$ are plane waves. As tracing back in time the outgoing rays of light near the event horizon and then to the sources of the ingoing light in the distance, the corresponding ingoing modes, $ {u}_{\bf k'}(x)$, will get distorted. The relation of the ingoing and outgoing rays of light can be obtained  by matching the metrics inside and outside the shell of a star during the collapse into a black hole. The corresponding coefficients  are then obtained as below
\begin{eqnarray*}
|\beta_{{\omega}}|^2=\frac{1}{e^{(2\pi/\kappa)  \omega}-1},\hspace{.1cm}{\rm and}\hspace{.3cm}
|\alpha_{{\omega, \omega'}}|^2=e^{(2\pi/\kappa)  \omega} |\beta_{{\omega, \omega'}}|^2.
\end{eqnarray*}
The Wightman function of the positive frequency can be expressed as
\begin{eqnarray*}
G^+(x,x')\equiv \langle 0,{\rm in}|\phi(x)\phi(x')|0,{\rm in}\rangle=\int\frac{d^n{\bf k}}{(2\pi)^3} {u}_{\bf k}(x) {u^*}_{\bf k}(x').
\end{eqnarray*}
where $|0,{\rm in}\rangle$ is the vacuum state for the ingoing wave functions. The Wightman function is used to calculate the detector response function, $\mathcal{F}(E)$, which indicates what particles experience along the trajectory of the detector, $x(t)$:
\begin{eqnarray}
\mathcal{F}(E)=\int^\infty_{-\infty} dt \int^\infty_{-\infty} dt' e^{-iE(t-t')}G^+(x(t),x(t')),\label{fofe}
\end{eqnarray}
thus the transition probability per unit time can be calculated. For a practical example, it is used to compute the depolarization of the particles coming from the synchrotron radiation for  the Sokolov-Ternov effect \cite{steffect} as mentioned in the introduction. Substitute the Bogoliubov transformation into eq. (\ref{bogo}), it then yields 
\begin{eqnarray*}
\hspace{0cm}G^+(x,x')&=&\int\frac{d^n{\bf k}}{(2\pi)^3} \left(|\alpha_{\omega}|^2\bar{u}_{\bf k}(x) \bar{u}^*_{\bf k}(x')+\alpha_{\omega}\beta^*_{\omega}\bar{u}_{\bf k}(x) \bar{u}_{\bf -k}(x')\right.\\
&&\hspace{3cm} \left.+\beta_{\omega}\alpha^*_{\omega}\bar{u}^*_{\bf -k}(x) \bar{u}^*_{\bf k}(x')+|\beta_{\omega}|^2\bar{u}^*_{\bf -k}(x) \bar{u}_{\bf -k}(x')\right),
\end{eqnarray*}
where only the final term in the above parenthesis survives from the $t$- and $t'$- integration in eq. (\ref{fofe}). The first term into eq. (\ref{fofe}) yields a delta function with a positive definite argument after integration over the new variables $t_-=t-t'$, $t_+=t+t'$, so its contribution vanishes as a result; the two terms in the middle also vanish after $t_+$-integration. 
On the other hand, from the imaginary-time approach the Wightman function can be constructed in a similar way to the photon fields. After summing over Matsubara frequencies and  analytic continuation to the real-time axis, the retarded Green function is 
\begin{eqnarray*}
\hspace{0cm}\langle 0|\phi(x)\phi(x')|0\rangle&=&\int \frac{d^n {\bf k}}{(2\pi)^3}\frac{1}{2\omega}\left(n_B(\omega)e^{i\omega(t_x-t_{x'})+i{\bf k\cdot (x-x')}}\right.\\
&&\hspace{3.2cm} \left.+(1+n_B(\omega))e^{-i\omega(t_x-t_{x'})-i{\bf k\cdot (x-x')}}\right).
\end{eqnarray*}
This is the exact form of the Green function for the scalar-field version, as compared with photon's case . As it is inserted into the detector response function, $\mathcal{F}(E)$, only the first term of the positive frequency in the parenthesis contributes, and the same result as that of the detector in a gravitational field can be acquired by corresponding the temperature of the vacuum bath to the strength of the gravity, $T=\frac{\kappa}{2\pi c k_B}$.
The final expressions of the detector response function from the two approaches are not presented here, since they are indeed identical to what is shown in the Chapter 3 of the first reference in ref. \cite{birrel}.


\section{Conclusion}
\label{conclusion}

Multiple vacuum effects have been reviewed and they all can be  understood in the approach  that starts from the imaginary-time partition function. A universal picture is granted to describe these phenomena caused by the vacuum of a finite temperature. From the previous work, it predicts consistent radiative corrections of the QED, including the anomalous magnetic dipole moments $g-2$, the Lamb shift, as well as the renormalization group equations. Besides the interactions between  photons and  matter, in this paper, the imaginary-time hamiltonians of the electromagnetic waves and fermions are also able to predict the Casimir effects and the van der Waals force and automatically provide regularization functions to avoid the divergence in the Matsubara's formalism. This means that, in the imaginary-time approach, the theory itself implies finite and consistent answers for the vacuum effects  from a thermodynamical perspective. For the Unruh effect, a realistic entity of a thermalized vacuum assumes the role of the thermal bath experienced by an accelerated observer. It is no longer just  an effective state of particles reaching a hypothetical thermal equilibrium; it could be  real. In a gravitational field, the thermodynamical mechanics of the Black holes is also shown to fit in the picture of the finite temperature vacuum. 
In short, the imaginary-time theory of vacuum explains various vacuum effects, which appear independent, in a universal way; more applications of this theory can be anticipated, such as the cosmological constant, and etc.. The Dewitt-Schwinger expansion \cite{birrel} provides a way to calculate the cosmological constant, unfortunately as the old problems that happens in the field theory, it is divergent. In the next work \cite{huang13c}, a finite answer for the dark energy density from the imaginary-time theory can be expected from two approaches, then the problems are if it can match the experimental data and if it is able to give a meaningful physics.  Therefore more tests and examinations both on the theory and experiments are necessary in the future.

 \appendix
\section{Lorentz gauge}
In calculating the Casimir effect and the van der Waals force, only the transverse polarizations of electromagnetic waves are taken into account. This is due to the Lorentz gauge is used to fix the gauge freedom. Separate the vector potential into positive and negative frequency parts, $A^\mu=A^\mu_{(+)}+A^\mu_{(-)}$. The gauge condition is applied, similar to what Gupta and Bleuler \cite{gupta} suggested, as 
\begin{eqnarray}
\left(i\frac{\partial}{\partial \tau}A^{0}_{(+)}+\nabla\cdot {\bf A}_{(+)}\right)|\Psi\rangle=0,\label{gupta}
\end{eqnarray}
where $|\Psi\rangle$ are the state vectors in Hilbert space. And we may obtain a similar expression for the adjoint condition for the negative frequency, $A^\mu_{(-)}$. To simplify the derivation without losing any generality, we may assume the propagating direction is $(i\omega_n,0,0,p^3)$ in the four-dimensional space-time.
The four polarization vectors can be defined as 
\begin{eqnarray*}
\epsilon_0^\mu=\frac{1}{|p|}(p^0,0,0,p^3),\hspace{.3cm}\epsilon_1^\mu=(0,1,0,0),\hspace{.3cm}
\epsilon_2^\mu=(0,0,1,0),\hspace{.3cm}
\epsilon_3^\mu=\frac{1}{|p|}(p^3,0,0,p^0),
\end{eqnarray*}
where $|p|=\sqrt{p^\mu p_\mu}$. It can be easily proved that the vectors are orthonormal and the summation of the scalar products $\sum^3_{\lambda=0} \epsilon^\mu_\lambda \epsilon_{\lambda,\mu}=\sum^3_{\lambda=0} g_{\lambda\lambda}$, where $g_{\lambda\lambda}$ is a matrix, not a tensor. From eq.  (\ref{gupta}), the gauge condition becomes for the annihilation operators of on-shell particles, $p^2=0$,
 \begin{eqnarray}
\left(a^0_{\omega_n,{\bf p}}+a^3_{\omega_n,{\bf p}}\right)|\Psi\rangle=0,
\end{eqnarray}
for any frequency and 3-momentum. It can be seen that the transverse part of the polarization vector $\epsilon^\mu_1$ and $\epsilon^\mu_2$ does not participate in the equation of gauge condition. Similarly to the conclusion in the real-time \cite{greiner}, the expectation value of the energy of the state, $|\Psi\rangle$, is only contributed by the  transverse part of the polarization.  



%
%

\end{document}